\documentclass[conference]{IEEEtran}
\IEEEoverridecommandlockouts
\usepackage{cite}
\usepackage{amsmath,amssymb,amsfonts}
\usepackage{algorithmic}
\usepackage{graphicx}
\usepackage{textcomp}
\usepackage{xcolor}
\usepackage{soul}

\usepackage[utf8]{inputenc} 
\usepackage[T1]{fontenc}    
\usepackage{hyperref}       
\usepackage{url}            
\usepackage{booktabs}       
\usepackage{amsfonts}       
\usepackage{nicefrac}       
\usepackage{microtype}  
\usepackage{lipsum}
\usepackage{array}
\usepackage{multirow}
\usepackage{wrapfig}
\usepackage{subfig}
\usepackage{textcomp}
\usepackage{fancyhdr}
\def\BibTeX{{\rm B\kern-.05em{\sc i\kern-.025em b}\kern-.08em
    T\kern-.1667em\lower.7ex\hbox{E}\kern-.125emX}}

\makeatletter
\newcommand{\linebreakand}{%
  \end{@IEEEauthorhalign}
  \hfill\mbox{}\par
  \mbox{}\hfill\begin{@IEEEauthorhalign}
}
\makeatother    

\begin{document}

\title{Accelerating SARS-CoV-2 low frequency variant calling on ultra deep sequencing datasets 
\thanks{Funded by the C3.ai Digital Transformation Institute COVID-19 award. MN was also funded by a fellowship from the National Library of Medicine Training Program in Biomedical Informatics and Data Science (T15LM007093, PI: Kavraki).}
}

\author{
\IEEEauthorblockN{Bryce Kille}
\IEEEauthorblockA{\textit{Department of Computer Science} \\
\textit{Rice University}, Houston, Texas\\
blk6@rice.edu}
\and
\IEEEauthorblockN{Yunxi Liu}
\IEEEauthorblockA{\textit{Department of Computer Science} \\
\textit{Rice University}, Houston, Texas\\
yl181@rice.edu}
\and
\IEEEauthorblockN{Nicolae Sapoval}
\IEEEauthorblockA{\textit{Department of Computer Science} \\
\textit{Rice University}, Houston, Texas\\
ns58@rice.edu}
\linebreakand
\IEEEauthorblockN{Michael Nute}
\IEEEauthorblockA{\textit{Department of Computer Science} \\
\textit{Rice University}, Houston, Texas\\
mn56@rice.edu}
\and
\IEEEauthorblockN{Lawrence Rauchwerger}
\IEEEauthorblockA{\textit{Department of Computer Science} \\
\textit{University of Illinois}, Urbana, Illinois\\
rwerger@illinois.edu}
\and
\IEEEauthorblockN{Nancy Amato}
\IEEEauthorblockA{\textit{Department of Computer Science} \\
\textit{University of Illinois}, Urbana, Illinois\\
namato@illinois.edu}
\linebreakand
\IEEEauthorblockN{Todd J. Treangen}
\IEEEauthorblockA{\textit{Department of Computer Science} \\
\textit{Rice University}, Houston, Texas\\
treangen@rice.edu}
}

\maketitle
\thispagestyle{fancy}
\begin{abstract}
With recent advances in sequencing technology it has become affordable and practical to sequence genomes to very high depth-of-coverage, allowing researchers to discover low-frequency variants in the genome. However, due to the errors in sequencing it is an active area of research to develop algorithms that can separate noise from the true variants. LoFreq is a state of the art algorithm for low-frequency variant detection but has a relatively long runtime compared to other tools. In addition to this, the interface for running in parallel could be simplified, allowing for multithreading as well as distributing jobs to a cluster. In this work we describe some specific contributions to LoFreq that remedy these issues.
\end{abstract}


\section{Introduction}

Cataloging viral mutations within a sample (intra-host variation) and across samples (inter-host variation) provides critical insights to understanding the dynamics of viral evolution during the COVID-19 pandemic~\cite{van2020emergence}. Single nucleotide variants (SNVs) can result in drastically different protein function and recognition; it is therefore desirable to be able to accurately and efficiently identify SNVs. For example, the SARS-CoV-2 virus was recently shown to have significant underlying diversity~\cite{sapoval2021hidden,karamitros2020sars}; additionally the mutations can change the fitness of the virus~\cite{plante2020spike} increasing it's transmission or pathogenicity potential~\cite{Davies2021}. 

Recent high throughput sequencing techniques enable researchers to generate read sets with deep coverage~\cite{reuter2015high}, allowing researchers to discover low frequency variants in the population. However, sequencing errors can masquerade as a low-frequency variant, and vice-versa, so distinguishing between the two is necessary~\cite{wang2019high}. A recent benchmarking study \cite{sandmann2017evaluating} found that on simulated HiSeq data, LoFreq~\cite{wilm2012lofreq} outperformed most variant calling tools, but suffered from long execution times. LoFreq operates by considering, at each position in the genome,
the probability that all mismatches to the reference genome are caused by sequencing error, although the specific probability calculation involved is computationally expensive and thus it does not scale well for deep sequencing datasets. 


In this work, we describe an improvement to the runtime of LoFreq by using an approximation to shortcut costly dynamic-programming operations when exact values of the probability distribution are not needed. We also add experimental support for OpenMP~\cite{dagum1998openmp} to enable more efficient parallel operation. 

\section{Methods}
Our approach to improve LoFreq is twofold. First, we implement an approximation heuristic which allows LoFreq to bypass computationally intensive exact probability calculations when not necessary. Second, we reorganize LoFreq to use OpenMP in lieu of a job-submission script for parallel operation. 

\subsection{Approximation}
As high-throughput sequencers read a DNA strand every position is assigned a quality score corresponding to the probability that the assigned nucleotide is incorrect. Thus, for a given pileup column (i.e. a position in the reference genome and the set of nucleotides at the corresponding position in each read that has been mapped to that location), the total number of sequencing errors is the sum of independent but \textit{not} identically distributed Bernoulli trials, also known as the Poisson binomial distribution.

\begin{figure*}[t!]%
\centering
\subfloat[]{%
\label{fig:first}%
\includegraphics[width=0.55\textwidth]{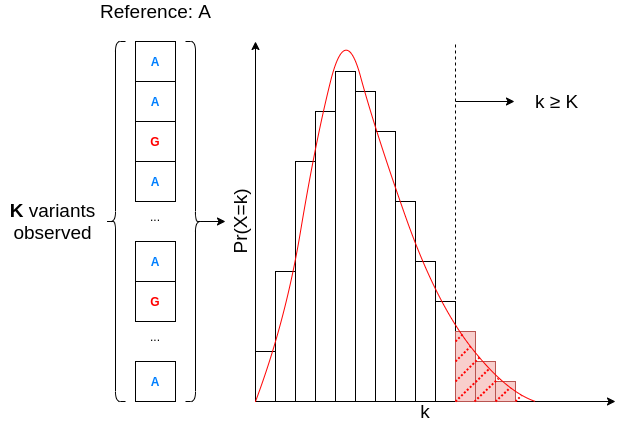}}%
\qquad
\subfloat[]{%
\label{fig:second}%
\includegraphics[width=0.35\textwidth]{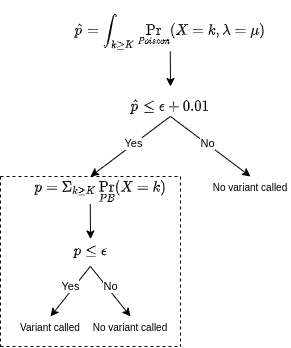}}%
\caption{\textbf{The continuous Poisson approximation (red line) to the Poisson binomial (bars) distribution}. \textbf{(a):} The test statistic for the Poisson approximation is the right tail integral (shaded) and for the Poisson binomial it is the right tail sum (red bars). \textbf{(b):} The workflow of the improved LoFreq algorithm. The original LoFreq workflow is denoted by the dotted box. Here $\Pr_{PB}$ denotes probability under the Poisson-Binomial distribution. We first compute the tail integral over the Poisson distribution to get an approximate $p$-value we denote as $\hat{p}$. If $\hat{p} > \epsilon + 0.01$, then we have high confidence that $p>\epsilon$ and therefore we do not call a variant. 
}
\label{fig:algo_fig}
\end{figure*}

More specifically, consider a single column with read depth $d$: let each read $i$ have a probability $p_i$ of having a sequencing error in this column (implied by the quality score), then the total error count is distributed as a Poisson Binomial with probabilities $\{p_i\}_{i=1}^d$. Although simple to parameterize, this distribution is computationally non-trivial, particularly for values of the cumulative distribution function (CDF) \cite{Biscarri2018poisson, hong2013computing}.    
Computing the probability of having \textit{at least} $K$ sequencing errors by chance requires the following recurrence relation:
$$
P_n(X{=}k)=P_{n-1}(X{=}k)(1{-}p_n) + P_{n-1}(X{=}k{-}1)p_n
$$ 
where $P_n(X{=}k)$ is the probability of observing $k$ errors in the first $n$ bases at the given position. Now, if $K$ reads in this column contain bases that differ from the reference genome, we can use the Poisson binomial CDF to calculate a $p$-value for the null hypothesis that the variants are due only to sequencing error: $p=\Sigma_{k\geq K} P_d(X{=}k)$. LoFreq operates in exactly this way, calling a variant in a given column if the $p$-value falls below a specified critical value (Figure \ref{fig:second}).

Unfortunately, this method of calculating the Poisson binomial takes $O(d^2)$ time; more recent algorithms may improve on this but remain complex ~\cite{Biscarri2018poisson, hong2013computing}. But for the vast majority of columns the $p$-value is far away from the decision threshold because either so few or so many variants are present, so bypassing the CDF calculation for these cases is a potential speedup.


We do this by using an $O(d)$ approximation to the Poisson binomial as a first pass filter: if the approximate $p$-value, $\hat{p}$, is sufficiently far from the critical value the exact CDF computation is skipped and no variant is called, otherwise the standard calculation is used (Figure~\ref{fig:second}). We have used the Poisson approximation where the mean is given by $\lambda=\sum_{i=1}^d p_i$ \cite{hodges1960poisson}, specifically using the GNU scientific library implementation~\cite{galassi2002gnu}. For our experiments, the significance threshold was left at the LoFreq default $\epsilon=0.05$ and the first pass filter required that $\hat{p} \ge \epsilon + 0.01$ in order to skip the expensive exact calculation.

\subsection{Parallelization}

The LoFreq algorithm operates by iterating through each pileup column checking for SNVs. The current parallel implementation uses an external script to parse the input files and and partition the columns equally and subsequently spawning an independent LoFreq process for partition. 
In an experimental branch of LoFreq, we implemented the same strategy using OpenMP rather than separate processes. This OpenMP version relies on a parallel \texttt{for} loop across chunks of columns, using an independent \texttt{.bam} file reader for each thread.

\section{Results}

\subsection{Performance Impact of the Poisson Approximation}
Experiments were ran using BAM files ranging from 1MB to 25GB generated from the FastQ data available in ~\cite{shotguncovid}. Both versions of LoFreq, the original and our improved version without the OpenMP functionality, were ran on an Intel Xeon Gold 6138 CPU with 64 threads for the benchmarks. Table~\ref{table:extime} shows the speedup observed with our approximation shortcut. The number of variants called by each version of LoFreq were identical for all five of the test data sets. 

\begin{figure*}[t!]%
\centering
\includegraphics[width=\textwidth]{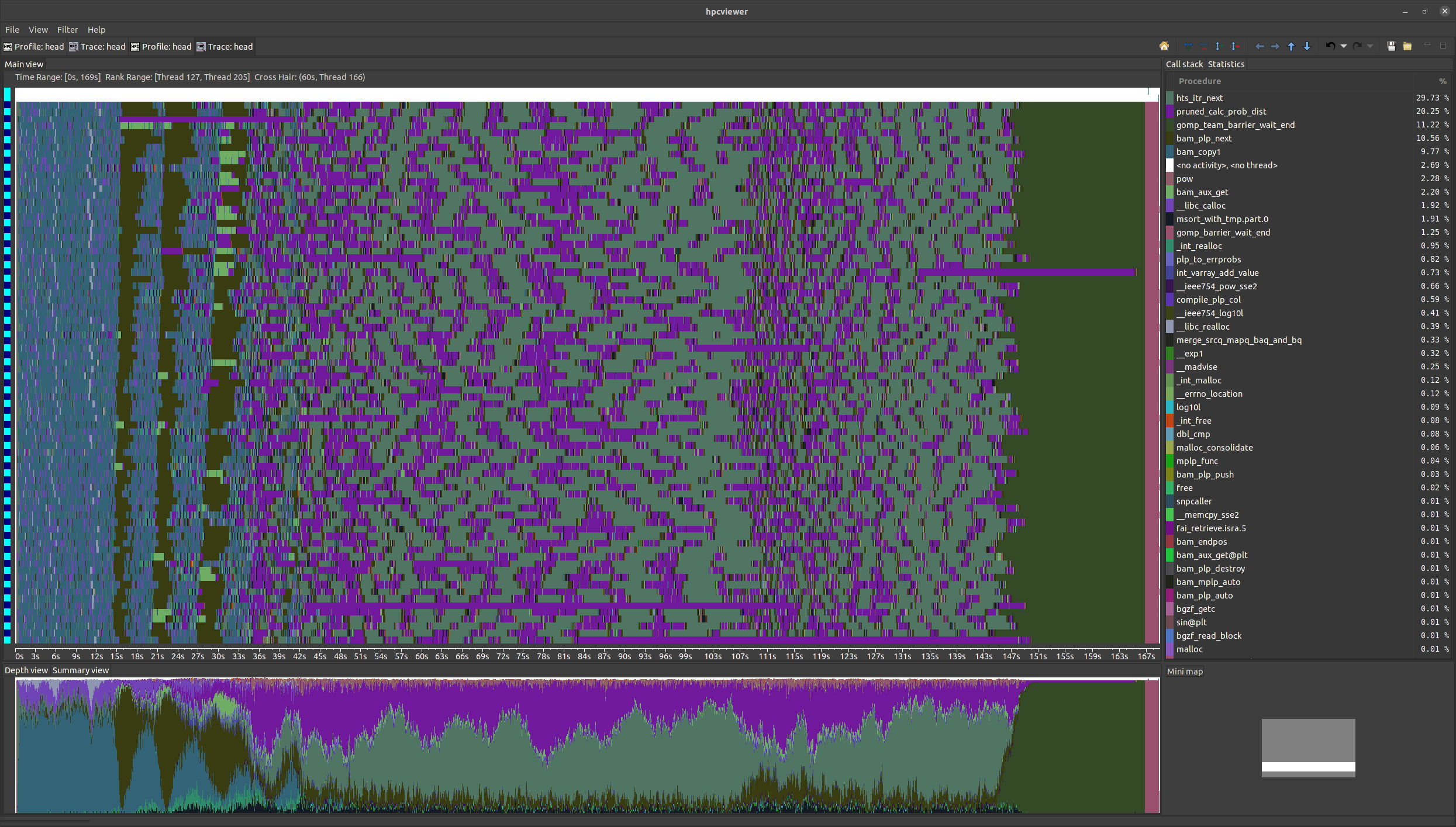}%
\caption{\textbf{ HPC-toolkit trace results.} X-axis is execution timeline and the Y-axis corresponds to the threads. The window at the bottom is the distribution of work across tasks. Pink is probability computation, teal is BAM file iteration, light blue at left is file decompression, dark green at right is the thread barrier. The image shows one thread causing a load imbalance due to a high-cost column.}
\label{fig:hpc_fig}
\end{figure*}

\begin{table}[t]
\centering
\caption{Execution times of the original and improved versions of LoFreq. In all cases the number of variants called was identical between versions. Note that while the true depth of the 25Gb file is likely around 5 million reads, LoFreq by default limits columns to 1 million.}\label{table:extime}
\begin{tabular}{|c|c|rr|c|}
    \toprule
    \multirow{2}[1]{*}{Input size} & \multirow{2}[1]{*}{Avg. depth} & \multicolumn{2}{c|}{Run Time} & \multirow{2}[1]{*}{Speed- up} \\
    
         & & Orig. & New   &      \\
    \midrule
    58M   &  1,000x & 52 (s) &  51 (s) & 1.0x \\
    237M  &  30,000x & 58 (m) &  26 (m) & 2.6x \\
    935M  &  100,000x & 14 (h) &  4 (h) & 3.3x \\
    2G    &  300,000x & 55 (h) &  12 (h) & 4.6x \\
    25G   &  1,000,000x & 415 (h) &  111 (h) & 3.7x \\
    \bottomrule
    \end{tabular}
\end{table}

\begin{figure}[t!]%
\centering
\includegraphics[width=0.45\textwidth]{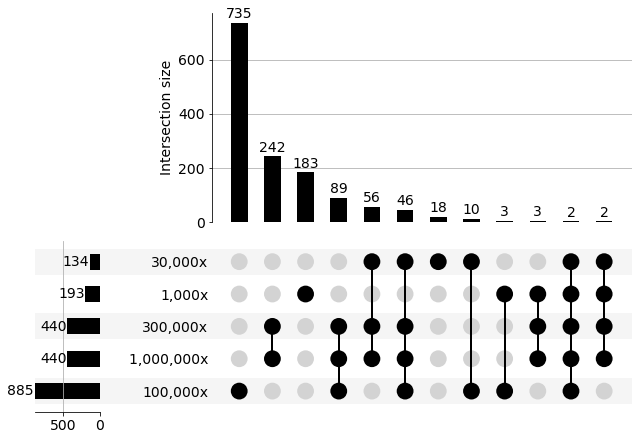}%
\caption{\textbf{Upset plot~\cite{lex2014upset} highlighting the shared low frequency variants across all five datasets.} Rows of upset plot indicate the depth-of-coverage per dataset, the columns indicate the intersection of shared single nucleotide variants across datasets. The bar plots located at the bottom left side represent the total number of SNVs identified per dataset.}
\label{fig:upset}
\end{figure}

\subsection{Profiling the OpenMP Implementation}
We utilized the HPC-toolkit~\cite{adhianto2010hpctoolkit} for visualizing the performance of the experimental OpenMP version of the improved LoFreq on a Knights Landing 2nd Generation Intel Xeon Phi processor with 128 threads. We observed in Figure~\ref{fig:hpc_fig}, as expected, that time spent coordinating threads is minimal and the process is trivially parallel over the columns in the input. We also notice that the time spent iterating over the \texttt{.bam} file is substantial. While our goal with OpenMP was to reduce load imbalance via dynamic scheduling, we see that encountering partitions with high concentrations of variants near the end still results in a significant imbalance. 


\subsection{SARS-CoV-2 single nucleotide variant analysis}
Finally, we performed an comparative analysis of SNVs identified across our COVID-19 datasets. We found from 134 (min) to 885 (max) SNVs in each dataset (Figure~\ref{fig:upset}). The two highest depth-of-coverage datasets, 300,000X and 1,000,000X, shared the most variants for any pair. The 100,000X dataset had the most unique SNVs at 735 total. Only two SNVs were found to be shared across all five datasets.



\section{Discussion}

In order to guarantee that the optimizations come at no cost to accuracy, we compared the output of the previous version of LoFreq to our improved version. Our improved version can only cause false negatives with respect to the original's variant calls, as we are using the approximation only to skip columns. Therefore, we only need to compare the count of variant calls in a sample across LoFreq versions. For all benchmarking datasets we observed the same number of variant calls in both versions of LoFreq.

Our results show that the approximation shortcut yields improved run-time over the original at no cost to accuracy in our samples. 
Since there is potential for our method to introduce false negatives into the results, we used an intentionally conservative threshold of $0.01$ above the critical value. No experimentation or fine-tuning was done to optimize this parameter or examine the feasibility of the same approach for very low values of $\hat{p}$, so that is a possible avenue for additional performance improvement. One approach could be to have the threshold vary according to read depth because the accuracy of the Poisson approximation increases at higher depth. 
As seen in the results, the approximation results in a 1-4x speedup in CPU time, particularly when run on large files with a low variant count. We also note that the approximation is more accurate when the error probabilities $p_i$ are higher, so a specific version of the algorithm could be optimized for high-error, long read sequencing data.

We observe that for input data with low read-depth this heuristic is actually ill-suited, and can even be both less accurate and slower. For one, as noted above the error in the Poisson approximation vanishes asymptotically as $d$ increases. Also, the existing version of LoFreq includes some conditions for early stopping that work especially well on shallow columns. To remedy this, our implementation only uses the approximation heuristic for columns with a read depth of at least 100. When read depth is below 100, the dynamic programming array used for the Poisson-Binomial fits inside the cache, which itself provides speedup comparable to the approximation. Importantly though, low-coverage sequencing input is inherently ill-suited to discovering variants present at low-frequency; for samples having depth below 100 throughout, a faster, less-sensitive SNV detector may be more appropriate.

The original LoFreq has not been optimized for cache performance, particularly on larger files. Since the computation of the Poisson binomial uses $O(d)$ memory, we quickly begin to spill over our shared cache when running in parallel files with depth $d > 1\mathrm{e}5$.
Our improved version of LoFreq has much better cache performance, with a cache miss rate below $15\%$ compared to over $70\%$ originally. Bypassing exact probability computations accounts for much of this as they repeatedly iterate over an array that does not fit in the cache. This also contributes to scalability in the new version since now, on average, only a small subset of running threads will need $O(d)$ memory at a given time. 

We were successfully able to replicate the behavior of the parallelization script in OpenMP. While this improvement does not reduce aggregate CPU time in it's current form on the experiments we have run, it does offer other advantages. The OpenMP implementation addresses a minor bug mentioned in a variant caller review article~\cite{sandmann2017evaluating} where the original implementation results in the output running through two stages of filtering when run in parallel: once for each individual process and then again on the combination of the variants from all of the processes. Unless set by the user, filter values are dynamically set during a LoFreq run, which causes the aforementioned filtering bug to produce inconsistent results. Our approach of using OpenMP to move all of the variant calling to the same process seems to remedy this problem. 

Additionally, the parallelization script from the original LoFreq approach could still be used to partition the input for submission to a cluster, making it possible to parallelize across both shared and distributed memory environments.
The OpenMP implementation also has the potential to avoid load imbalances that were possible previously by using smaller partitions towards the end of the run.

\section{Conclusion}
These modifications to two small parts of the LoFreq source code have improved an existing, widely used variant calling software and would not have been possible without a flexible and well-documented code base, which along with their spirit of collaboration is a credit to the developers. The effect of it is the continuous improvement of a single software, a welcome change from the more common pattern where bioinformatics tools proliferate every time a minor modification is made to an algorithm. 

%

    Our heuristic improvements to LoFreq have been merged to the main repository, available at \url{https://github.com/CSB5/lofreq}. The experimental OpenMP version is available at \url{https://gitlab.com/treangenlab/lofreq/-/tree/openmp/}

\section{Acknowledgments}
We'd like to thank the authors of LoFreq, Andreas Wilm and Niranjan Nagarajan, for their help and advice as well as merging our improvements into the LoFreq repository. We'd also like to thank John Mellor-Crummey for assistance with HPC-Toolkit and access to additional computing resources.
\label{sec:headings}

\bibliographystyle{unsrt}  

\bibliography{references}

\end{document}